\documentclass{ws-procs975x65}            

\usepackage[utf8]{inputenc}

\usepackage{amsmath}

\usepackage{amsfonts}
\usepackage{mathrsfs}

\usepackage{units}

\usepackage{placeins}

\usepackage{verbatim}

\usepackage{accents}

\newcommand{\mit}{\mathit}
\newcommand{\mrm}{\mathrm}

\makeatletter

\let\old@dmathbeg\[
\let\old@dmathend\]

\newcommand{\rovnec}[1]{\old@dmathbeg#1\old@dmathend}
\newcommand{\rovcis}[2]{\begin{equation}#1\label{#2}\end{equation}}
\newcommand{\drovcis}[2]{\begin{equation}\begin{split}#1\end{split}\label{#2}\end{equation}}
\newcommand{\drovnec}[1]{\begin{equation*}\begin{split}#1\end{split}\end{equation*}} 
\newcommand{\provcis}[1]{\begin{align}#1\end{align}}
\newcommand{\provnec}[1]{\begin{align*}#1\end{align*}}

\newcommand{\rov}{\@ifstar\rovnec\rovcis}
\newcommand{\drov}{\@ifstar\drovnec\drovcis}
\newcommand{\prov}{\@ifstar\provnec\provcis}

\newcommand{\vast}{\bBigg@{4}}
\newcommand{\Vast}{\bBigg@{5}}

\makeatother

\DeclareMathOperator{\sgn}{sgn}
\DeclareMathOperator{\diffbold}{\mathbf{d}}
\newcommand{\bd}{\diffbold\!}

\newcommand{\msc}{\mathscr}
\newcommand{\mbs}{\boldsymbol}

\newcommand{\iDelta}{{\mit\Delta}}
\newcommand{\iSigma}{{\mit\Sigma}}
\newcommand{\iLambda}{{\mit\Lambda}}
\newcommand{\iXi}{{\mit\Xi}}

\DeclareFontEncoding{LGR}{}{}

\DeclareMathAlphabet{\mgr}{LGR}{cmr}{m}{n}
\newcommand{\rpi}{\mgr{p}}

\renewcommand{\[}{\left[}
\renewcommand{\]}{\right]}

\newcommand{\f}{\!\left}

\newcommand{\zrov}{{}\\{}}

\newcommand{\res}[2]{\left.#1\right|_{#2}}
\newcommand{\lbl}{\label}
\newcommand{\rvt}{\ .}
\newcommand{\rvc}{\ ,}
\newcommand{\qt}[1]{``#1''}

\renewcommand{\(}{\left(}
\renewcommand{\)}{\right)}

\begin{document}

\title{Extremal black holes in strong magnetic fields: near-horizon geometries and Meissner effect}

\author{Filip Hejda} 

\address{Centro Multidisciplinar de Astrofísica -- CENTRA, Departamento de
Física, Instituto Superior Técnico -- IST, Universidade de Lisboa -- UL, Avenida Rovisco Pais 1, 1049-001 Lisboa, Portugal}

\address{Institute of Theoretical Physics, Faculty of Mathematics and Physics,
Charles University in Prague,
V Holešovičkách 2, 180\,00 Prague 8, Czech Republic}

\author{Jiří Bičák}

\address{Institute of Theoretical Physics, Faculty of Mathematics and Physics,
Charles University in Prague,
V Holešovičkách 2, 180\,00 Prague 8, Czech Republic}

\begin{abstract}
For extremal black holes, one can construct simpler, limiting spacetimes that describe the geometry near degenerate horizons. Since these spacetimes are known to have enhanced symmetry, the limiting objects coincide for different solutions. We show that this occurs for strongly magnetised Kerr-Newman solution, and how this is related to the Meissner effect of expulsion of magnetic fields from extremal black holes.
\end{abstract}


\bodymatter

\section{Magnetised Kerr-Newman Black Holes}

The general MKN (magnetised Kerr-Newman) black hole has the metric 
\rov{\mbs{g}=\left|\iLambda\right|^2\iSigma\[-\frac{\iDelta}{\msc{A}}\bd t^2+\frac{\bd r^2}{\iDelta}+\bd\vartheta^2\]+\frac{\msc A}{\iSigma\left|\iLambda\right|^2}\sin^2\vartheta\(\bd\varphi-\omega\bd t\)^2}{ewmkn} with $\iDelta=r^2-2Mr+a^2+Q^2$, $\iSigma=r^2+a^2\cos^2\vartheta$, $\msc A=\(r^2+a^2\)^2-\iDelta a^2\sin^2\vartheta$,
 and function $\iLambda$ (involving magnetic field parameter $B$), which reads \drov{\iLambda=1&+\frac{1}{4}B^2\(\frac{\msc A+a^2Q^2\(1+\cos^2\vartheta\)}{\iSigma}\sin^2\vartheta+Q^2\cos^2\vartheta\)+\zrov&+\frac{BQ}{\iSigma}\[ar\sin^2\vartheta-\mrm{i}\(r^2+a^2\)\cos\vartheta\]-\zrov&-\frac{\mrm{i}}{2}B^2a\cos\vartheta\[M\(3-\cos^2\vartheta\)+\frac{Ma^2\sin^2\vartheta-Q^2r}{\iSigma}\sin^2\vartheta\]\rvt}{lammkn}
The most complete picture of a general MKN black hole was recently given by Gibbons, Mujtaba and Pope \cite{Pope}, where the expressions for the dragging potential $\omega$ and electromagnetic potential $A_\iota$ can be found. 

\section{Near-Horizon Limit}

Metric of a stationary, axially symmetrical extremal black hole can be put into the form
\rov{\mbs g=-\(r-r_0\)^2\tilde N^2\bd t^2+g_{\varphi\varphi}\(\bd\varphi-\omega\bd t\)^2+\frac{\tilde g_{rr}}{\(r-r_0\)^2}\bd r^2+g_{\vartheta\vartheta}\bd\vartheta^2\rvc}{axst2}
where the degenerate horizon is located at $r=r_0$ and $\tilde N, \tilde g_{rr}$ are regular and non-vanishing there.

For extremal black holes, there exists yet another infinity apart from \qt{standard} spatial infinity, since proper radial distance between two points along $t=\mrm{const.}$ diverges, if one of the points approaches $r_0$. To extract the geometry of this \qt{throat} region involving $r=r_0$, we introduce coordinate transformation
\enlargethispage{\baselineskip}
\prov{r&=r_0+p\chi\rvc&t&=\frac{\tau}{p}\rvc\label{rchittau}}
and take the limit $p\to0$. We also have to get rid of the horizon values of the dragging and electrostatic potential by \qt{rewinding} the azimuth and changing the gauge\cite{Comp}
\prov{\varphi&=\psi+\frac{\omega_\mrm{H}}{p}\tau\rvc&\mbs A&\to\mbs A-\(\res{A_t}{r_0}+\omega_\mrm{H}\res{A_\varphi}{r_0}\)\frac{\bd\tau}{p}\equiv\mbs A+\frac{\phi_\mrm{H}}{p}\bd\tau\rvt}
If we apply the \qt{recipe} to the Kerr-Newman solution, the resulting metric \cite{Carter73} is 
\drov{\mbs g=\[Q^2+a^2\(1+\cos^2\vartheta\)\]\(-\frac{\chi^2}{\(Q^2+2a^2\)^2}\bd\tau^2+\frac{\bd\chi^2}{\chi^2}+\bd\vartheta^2\)+\zrov+\frac{\(Q^2+2a^2\)^2}{Q^2+a^2\(1+\cos^2\vartheta\)}\sin^2\vartheta\(\bd\psi+\frac{2a\sqrt{Q^2+a^2}\chi}{\(Q^2+2a^2\)^2}\bd\tau\)^2\rvc}{knnh}
whereas the electromagnetic potential \cite{Comp} reads 
\rov{\mbs A=\frac{Q}{Q^2+a^2\(1+\cos^2\vartheta\)}\(\frac{Q^2+a^2\sin^2\vartheta}{Q^2+2a^2}\chi\bd\tau+a\sqrt{Q^2+a^2}\sin^2\vartheta\bd\psi\)\rvt}{aknnh}
It can be shown that the components of $A_\mu$ in the near-horizon limit satisfy
\prov{A_\tau&=\res{\(\tilde N\sqrt{\tilde g_{rr}}F_{(r)(t)}\)}{r_0}\chi\equiv\tilde A_\tau\f(\vartheta\)\chi\rvc&
A_\psi\f(\vartheta\)&=\frac{1}{\tilde\omega}\(\tilde A_\tau\f(0\)-\tilde A_\tau\f(\vartheta\)\)\rvc}
where $\tilde\omega$ is the radial derivative of $\omega$ at $r_0$. It is useful for solutions generated by Harrison transformation \cite{Ernst76}, like for MKN black holes, to express $A_\alpha$ in terms of $F_{(r)(t)}$.

MKN solution is described by four parameters $M,Q,a,B$. We are interested in extremal cases, so $M^2=Q^2+a^2$. Interestingly, the geometry of extremal MKN black holes in the near-horizon limit coincides with that of Kerr-Newman black holes. However, the near-horizon geometry of a corresponding Kerr-Newman black hole is not described by parameters $Q, a$ of the MKN black hole, rather by \emph{effective parameters} $\hat M,\hat Q,\hat a$ given as follows:
\prov{\hat M&=\sqrt{Q^2+a^2}\(1+\frac{1}{4}B^2Q^2+B^2a^2\)+BQa\rvc\lbl{meff}\\\hat a&=a\(1-\frac{3}{4}B^2Q^2-B^2a^2\)-BQ\sqrt{Q^2+a^2}\rvc\lbl{aeff}\\\hat Q&=Q\(1-\frac{1}{4}B^2Q^2\)+2Ba\sqrt{Q^2+a^2}\rvt\lbl{qeff}}
These again satisfy the condition ${\hat M}^2={\hat Q}^2+{\hat a}^2$.\footnote{$\hat M$ is positive, so that we can use it in place of $\sqrt{{\hat Q}^2+{\hat a}^2}$.} When we substitute $\hat Q, \hat a$ in place of $Q,a$ in \eqref{knnh} and \eqref{aknnh}, we obtain the near-horizon metric and potential of the \qt{original} MKN black hole. However, to see this, we have to employ rescaled coordinates $\psi\to\iXi^{-1}\psi,\tau\to\iXi\tau$, with 
the constant factor
\rov{\iXi=\frac{1}{1+\frac{3}{2}B^2Q^2+2B^3Qa\sqrt{Q^2+a^2}+B^4\(\frac{1}{16}Q^4+Q^2a^2+a^4\)}\rvt}{}
Since it holds 
\rov{\frac{Q^2+2a^2}{\iXi}={\hat Q}^2+2{\hat a}^2\rvc}{}
the metric can be written in the form
\drov{\mbs g=\(\hat M^2+\hat a^2\cos^2\vartheta\)\(-\frac{\chi^2}{\(Q^2+2a^2\)^2}\bd\tau^2+\frac{\bd\chi^2}{\chi^2}+\bd\vartheta^2\)+\zrov+\frac{\(Q^2+2a^2\)^2\sin^2\vartheta}{\hat M^2+\hat a^2\cos^2\vartheta}\(\bd\psi+\frac{2\hat M\hat a\chi}{\(Q^2+2a^2\)^2}\bd\tau\)^2\rvc}{mknnheff}
whereas the electromagnetic potential reads
\rov{\mbs A=\frac{\hat Q}{\hat M^2+\hat a^2\cos^2\vartheta}\(\frac{\hat Q^2+\hat a^2\sin^2\vartheta}{Q^2+2a^2}\chi\bd\tau+\hat M\hat a\sin^2\vartheta\iXi\bd\psi\)\rvt}{amknnheff}
Final metric \eqref{mknnheff} and potential \eqref{amknnheff} coincide with those obtained by the application of the near-horizon limit to the original MKN solution. 

The idea that the near-horizon limits of the Kerr-Newman and MKN solutions are equivalent, up to a non-trivial reparametrisation and some rescaling, dates back to 2013 to theses of Hejda\cite{dipl} and Hunt\cite{Hunt}. It was fully developed in 2015\cite{Article1, Booth}, where it was further discussed that this fact can be also inferred using certain uniqueness theorems like the one of Lewandowski and Pawlowski\cite{LP}. 

Furthermore, it is remarkable that when we define the angular momentum of extremal MKN black holes by $\hat J=\hat a\hat M$,
i.e. in analogy with the standard case without magnetic field, we find that the result coincides precisely with the thermodynamic angular momentum given by Gibbons, Pang and Pope \cite{GibbonsPope2} in formula (5.11), when it is restricted to the extremal case. We already know that, in properly rescaled coordinates, the near-horizon limit of any extremal MKN black hole can be described just by effective parameters $\hat Q, \hat a$. We can express $\hat a$ using $\hat Q,\hat J$ as follows:
\rov{\hat a=\frac{\sgn\hat J}{\sqrt{2}}\sqrt{\sqrt{{\hat Q}^4+4{\hat J}^2}-{\hat Q}^2}\rvt}{}
Since $\hat Q$ is a physical charge of the black hole and $\hat J$ is derived by Gibbons, Pang and Pope \cite{GibbonsPope2} as a meaningful angular momentum, we may conclude that the near-horizon limit of any extremal MKN black hole can be described in terms of thermodynamic charges of the black hole.

\section{Parameter Space}

Understanding of near-horizon geometry of extremal MKN black holes can have multiple applications. For example, one can use it in connection with concepts like holographic duality\cite{Astorino}. Another question is the structure of the parameter space of extremal MKN black holes. In our analysis\cite{Article1} we ignore mass, which is just a scale and use dimensionless parameters $BM$ and $\gamma_\mrm{KN}$, which represents \qt{mixing} of \emph{bare} rotation parameter and charge: $a=M\cos\gamma_\mrm{KN},Q=M\sin\gamma_\mrm{KN}$.

As we discussed before, the near-horizon geometry of extremal MKN black holes is fully described by effective parameters $\hat Q, \hat a$. MKN black holes with the same ratios between these parameters will have the same near-horizon properties. To visualise this, we can introduce another mixing angle by
\rov{\hat Q=\hat M\sin\hat\gamma_\mrm{KN}\rvt}{mknfol0}
Using expressions \eqref{qeff} and \eqref{meff} for $\hat Q$ and $\hat M=\sqrt{{\hat Q}^2+{\hat a}^2}$, we can solve the relation with respect to $B$
\rov{B=\frac{4a\sqrt{Q^2+a^2}-2Qa\sin\hat\gamma_\mrm{KN}\mp2\(Q^2+2a^2\)\cos\hat\gamma_\mrm{KN}}{Q^3+\(Q^2+4a^2\)\sqrt{Q^2+a^2}\sin\hat\gamma_\mrm{KN}}}{}
and then express the solution in terms of dimensionless $\gamma_\mrm{KN}$ and $BM$ as follows\footnote{The $\mp$ sign is not necessary if we extend the interval for $\hat\gamma_\mrm{KN}$ to  $\(-\rpi,\rpi\]$. Then the minus sign in front of the term with $\cos\hat\gamma_\mrm{KN}$ gurantees that $\hat\gamma_\mrm{KN}=0$ implies $\hat a=\hat M$.}: 
\rov{BM=\frac{4\cos\gamma_\mrm{KN}-\sin2\gamma_\mrm{KN}\sin\hat\gamma_\mrm{KN}\mp2\(1+\cos^2\gamma_\mrm{KN}\)\cos\hat\gamma_\mrm{KN}}{\sin^3\gamma_\mrm{KN}+\(1+3\cos^2\gamma_\mrm{KN}\)\sin\hat\gamma_\mrm{KN}}\rvt}{}
This formula describes the distribution of equivalent near-horizon geometries viewed as curves in the parameter space of $\gamma_\mrm{KN}, BM$. Such curves for various values of $\hat\gamma_\mrm{KN}$ are plotted in Figure \ref{MKNps}. Four of them, given by four special values of $\hat\gamma_\mrm{KN}$ (multiplies of $\nicefrac{\rpi}{2}$) represent special cases $\hat a=0$ or $\hat Q=0$.

\begin{figure}
\centering
\input{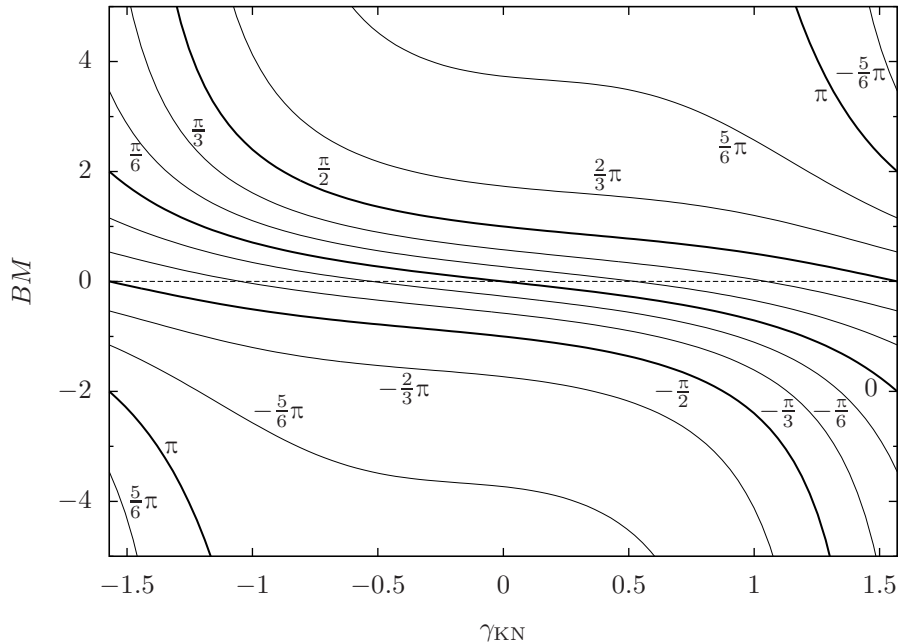}
\caption{Near-horizon geometries of extremal MKN black holes illustrated in the dimensionless parameter space $\(\gamma_\mrm{KN},BM\)$. Each curve describes geometries with fixed parameter $\hat\gamma_\mrm{KN}$, i.e. with fixed ratios $\nicefrac{\hat Q}{\hat M}$ and $\nicefrac{\hat a}{\hat M}$ (see equation \eqref{mknfol0}). Various regions in the plane correspond to different signs of parameters $\hat a,\hat Q$, respectively. The boundaries between the regions are indicated by thick lines.}
\lbl{MKNps}
\end{figure}

\section{Meissner Effect}

It is carefully reviewed by Bičák, Karas and Ledvinka \cite{BiKaLe} that external stationary, axisymmetric \emph{test} magnetic fields are expelled from the degenerate horizons. This is referred to as a black-hole Meissner effect. It is natural to ask whether such effect is present even in the strong field regime.
Karas and Vokrouhlický \cite{KarVok91} discussed the total magnetic flux across the upper hemisphere of the horizon of extremal MKN black holes and found that in two special cases corresponding to our $\hat a=0$ and $\hat Q=0$ the flux vanishes. 

In these cases our results support their conclusion. Indeed, the magnetic field is encoded in the component $A_\psi$ of the electromagnetic potential, given in \eqref{amknnheff}, which is proportional to the product $\hat Q\hat a$. 

For a general extremal MKN black hole the magnetic flux through upper hemisphere of the horizon does not vanish. It can be expressed as follows:
\rov{\msc F_\mrm{H}=2\rpi\frac{\res{A_\psi}{\vartheta=\frac{\rpi}{2}}}{\iXi}=2\rpi\frac{\hat Q\hat a}{\hat M}=\frac{4\rpi\hat Q\hat J}{{\hat Q}^2+\sqrt{{\hat Q}^4+4{\hat J}^2}}\rvt}{}
Since the structure of the azimuthal component of the electromagnetic potential in the near-horizon limit is identical (up to the rescaling) to the one of the Kerr-Newman black hole, the flux can be expressed using the Kerr-Newman-like effective parameters. As we stated above, these parameters can be related to thermodynamic charges of the black hole \cite{GibbonsPope2}. Therefore, we may conclude that the magnetic flux is intrinsic to the black hole configuration and there is no flux caused directly by the external magnetic field.

\section*{Acknowledgement}

The work was supported by research grants: SVV No. 260211 and 265301, GAČR No. 14-37086 G and GAUK No. 606412. F.H. would also like to thank Tomáš Benedikt for his ready technical support.

\end{document}